\documentclass[runningheads]{llncs}
\usepackage[utf8]{inputenc}
\usepackage{textcomp}
\usepackage{graphicx,verbatim}
\usepackage{hyperref}
\usepackage{cite}
\usepackage{amsmath}
\usepackage{amssymb}
\usepackage{authblk}
\usepackage{bbding}
\usepackage{graphicx}
\usepackage{booktabs}
\usepackage{graphicx}
\DeclareMathOperator*{\argmin}{arg\,min}

\usepackage[normalem]{ulem}
\usepackage{multirow}
\usepackage{amsmath} 
\usepackage{bm}      
\usepackage{multirow}
\usepackage[table,xcdraw]{xcolor}
\usepackage{colortbl}
\usepackage[normalem]{ulem}
\useunder{\uline}{\ul}{}
\usepackage[ruled,vlined]{algorithm2e}
\usepackage{algpseudocode} 
\begin{document}
\title{A Composite Alignment-Aware Framework for Myocardial Lesion Segmentation in Multi-sequence CMR Images}
\titlerunning{Composite Alignment-Aware Framework}
%

\author{Yifan Gao \inst{1,2}, 
Shaohao Rui\inst{2,3,5},  
Haoyang Su\inst{2,4,5}, 
Jinyi Xiang\inst{6}, 
Lianming Wu\inst{6}, 
Xiaosong Wang\inst{2,5}\textsuperscript{\Envelope}}  

\authorrunning{Y. Gao et al.}

\institute{
School of Biomedical Engineering (Suzhou), Division of Life Science and Medicine, University of Science and Technology of China, Hefei, China \and
Shanghai Innovation Institute, Shanghai, China \and
Shanghai Jiao Tong University, Shanghai, China \and
Fudan University, Shanghai, China \and
Shanghai Artificial Intelligence Laboratory, Shanghai, China \and
Department of Radiology, Renji Hospital, Shanghai Jiao Tong University, Shanghai, China \\ }

\maketitle              

\begingroup
\renewcommand\thefootnote{\textsuperscript{\Envelope}}
\footnotetext[2]{Corresponding author}
\endgroup
\begin{abstract}
Accurate segmentation of myocardial lesions from \ multi-sequence cardiac magnetic resonance imaging is essential for cardiac disease diagnosis and treatment planning. However, achieving optimal feature correspondence is challenging due to intensity variations across modalities and spatial misalignment caused by inconsistent slice acquisition protocols. We propose CAA-Seg, a composite alignment-aware framework that addresses these challenges through a two-stage approach. First, we introduce a selective slice alignment method that dynamically identifies and aligns anatomically corresponding slice pairs while excluding mismatched sections, ensuring reliable spatial correspondence between sequences. Second, we develop a hierarchical alignment network that processes multi-sequence features at different semantic levels, i.e., local deformation correction modules address geometric variations in low-level features, while global semantic fusion blocks enable semantic fusion at high levels where intensity discrepancies diminish. We validate our method on a large-scale dataset comprising 397 patients. Experimental results show that our proposed CAA-Seg achieves superior performance on most evaluation metrics, with particularly strong results in myocardial infarction segmentation, representing a substantial 5.54\% improvement over state-of-the-art approaches. The code is available at https://github.com/yifangao112/CAA-Seg.

\end{abstract}
\section{Introduction}

Cardiovascular diseases remain a leading global health burden, with myocardial pathologies such as infarction and edema requiring precise quantification for effective clinical management \cite{vaduganathan2022global,rui2025cardiocot}. While manual segmentation of these pathological regions is time-consuming and subject to significant inter-observer variability \cite{gao2024desam,gao2024mba}, automatic segmentation presents substantial challenges due to the small size and subtle appearance of infarction regions. Multi-sequence cardiac magnetic resonance (CMR) imaging has emerged as a cornerstone for comprehensive myocardial assessment, where late gadolinium enhancement (LGE) delineates lesion with high contrast \cite{unger2025prognostic}, while complementary T1 and T2 mapping sequences provide additional pathological indicators for both infarction and edema \cite{liu2022cardiac,warnica2022clinical}. 

\begin{figure}[t!]
	\includegraphics[width=\textwidth]{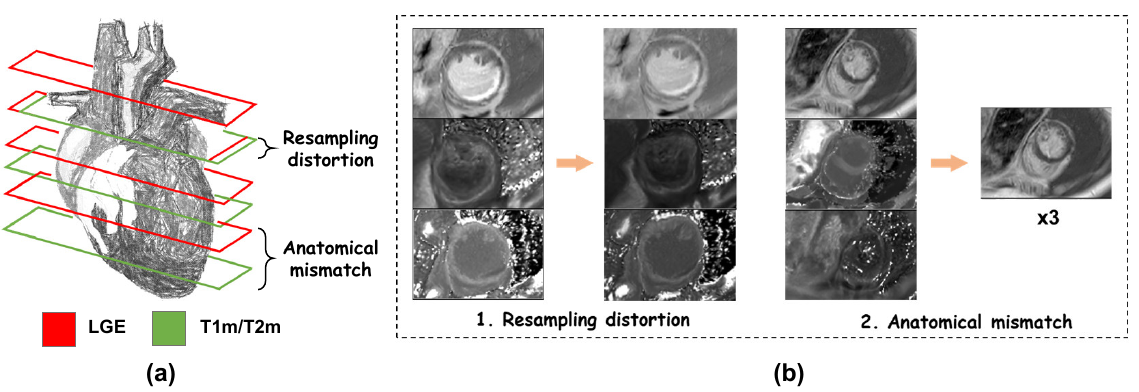}
	\centering
	\caption{(a) Visualization of misalignment challenges in multi-sequence CMR. (b.1) \textbf{Resampling distortion:} Aggressive resampling of T1/T2 mapping (3 to 8 slices) introduces interpolation artifacts and structural distortion. (b.2) \textbf{Anatomical mismatch:} LGE and T1/T2 mapping slices were acquired at different anatomical positions, causing misalignment. These challenges motivate our selective slice alignment strategy that preserves only reliable correspondences while avoiding forced matching of incompatible slices. Our slice-level alignment significantly reduces resampling-induced misregistration. Additionally, we use only duplicated LGE slices to avoid anatomical mismatch.} 
	\label{fig0}
\end{figure}

Leveraging these imaging sequences offers promising opportunities to enhance segmentation performance by combining distinct pathological markers from each modality \cite{gao2023anatomy,dai2021transmed}. However, the integration of these complementary sequences faces two significant technical barriers, as shown in Fig. \ref{fig0}. First, the substantial difference in slice sampling between sequences creates severe anatomical mismatches. LGE typically employs 6-8 slices to capture localized pathologies, whereas T1/T2 mapping sequences use only 2-3 slices for rapid quantification. When these slices are acquired at different anatomical positions, direct registration becomes anatomically inconsistent and clinically unreliable (Fig. 1b.2). Second, attempting to resolve this mismatch through aggressive resampling introduces significant artifacts. Resampling T1/T2 mapping from 3 to 8 slices forces interpolation across large gaps, creating artificial structural continuity and distorting the very pathological features we aim to segment (Fig. 1b.1). Thus, the challenge lies in establishing both inter-sequence semantic correspondence and intra-sequence spatial consistency across these heterogeneously sampled data streams.

Existing methods struggle to address these challenges effectively. Conventional approaches predominantly focus on single-modality analysis \cite{yue2019cardiac,yu2021cardiac,li2022medical}, particularly LGE, sacrificing complementary pathological information from other sequences. While recent multi-modal fusion attempts \cite{jiang2020max,wang2022awsnet,qiu2023myops,xu2025multi} demonstrate promise on small datasets (25 patients) \cite{li2023myops} with pre-aligned inputs, such idealized conditions rarely reflect the misalignment prevalent in routine clinical acquisitions. This discrepancy forces networks to process anatomically inconsistent features, particularly in mid-ventricular regions where interpolation artifacts dominate.

To address these challenges, we propose CAA-Seg, a composite alignment-aware framework for myocardial lesion segmentation in multi-sequence CMR images. The framework consists of two sequential stages: selective slice alignment (SSA) and hierarchical feature fusion. First, we propose a selective slice alignment method that handles \textit{resampling distortion} and \textit{anatomical mismatch} through dynamic slice matching. Instead of forcing alignment between all slices, SSA employs sliding-window optimization to identify anatomically corresponding slice pairs while excluding mismatched regions. This generates aligned multi-sequence regions for lesion fusion and single-sequence regions for standard segmentation, which are processed through task-specific channels.
Second, we develop a hierarchical alignment network (HA-Net) that addresses multi-sequence feature fusion at different semantic levels. At low-level features where local geometric variations dominate, we employ deformable convolutions and pixel-wise modulation to correct spatial misalignments. High-level features, which encode pathological patterns with reduced intensity variations, are aligned through cascaded cross-attention modules to capture semantic relationships between LGE and mapping sequences. A task-aware controller at the bottleneck level adaptively modulates features based on input characteristics, ensuring effective fusion of multi-sequence information.

Our comprehensive validation on a large-scale CMR dataset demonstrates significant improvements over state-of-the-art methods. CAA-Seg achieves 78.09\% Dice score for myocardium segmentation and 65.49\% for myocardial edema delineation. Notably, for myocardial infarction segmentation, our method achieves 51.11\% Dice score, outperforming the second-best approach by a substantial margin of 5.54\%. The main technical contributions are:

\begin{enumerate}
  \item We propose a novel framework for multi-sequence CMR segmentation that effectively integrates heterogeneous cardiac imaging sequences through composite alignment.
  \item We propose a selective slice alignment approach that establishes reliable inter-sequence correspondence through dynamic slice matching and mutual information optimization.
  \item We develop a hierarchical alignment network that combines local deformation correction and global semantic fusion for effective multi-sequence feature fusion.
\end{enumerate}

\section{Method}
We propose a two-stage framework for myocardial lesion segmentation in multi-sequence CMR images. The first stage employs selective slice alignment to establish anatomical correspondence between LGE and mapping sequences through dynamic slice matching. The aligned sequences then serve as input to a hierarchical alignment network (HA-Net), which performs multi-level feature fusion while addressing residual spatial and intensity variations for accurate lesion segmentation.

\subsection{Selective Slice Alignment}
Multi-sequence CMR acquisition results in varying slice counts and positions across sequences, making direct registration between LGE and mapping sequences challenging. Rather than forcing alignment between all slices, which can introduce artifacts and misleading information, we propose the SSA approach that dynamically identifies and aligns only the most reliable slice pairs.

Given a series of 2D slices, let $I_f^k \in \mathbb{R}^{H \times W}$ denote the $k$-th fixed LGE slice and $I_m^j \in \mathbb{R}^{H \times W}$ represent the $j$-th moving slice from either T1m or T2m sequence. We formulate the slice-wise registration as an optimization problem:

\begin{equation}
        \phi_k^* = \arg\min_{\phi_k} \underbrace{\mathcal{L}_{MMI}(I_f^k, I_m^j \circ \phi_k)}_{\text{Multi-sequence similarity}} + \lambda \underbrace{\mathcal{R}(\phi_k)}_{\text{Regularization}},\quad k \in [1,K]
\end{equation}

\noindent where $K$ is the total number of slices. To ensure robust multi-modal alignment, we adopt Mattes mutual information (MMI) as the similarity metric:

\begin{equation}
\mathcal{L}_{MMI}(I_f^k, I_m^j \circ \phi_k) = -\sum_{x,y} p_k(x,y) \log\left(\frac{p_k(x,y)}{p_f^k(x)p_m^j(y)}\right)
\end{equation}

\noindent where $p_k(x,y)$ represents the joint intensity distribution. Our transformation model accommodates both global and local deformations through a hierarchical composition:

\begin{equation}
\phi_k = \phi_{k,rigid} \circ \phi_{k,affine} \circ \phi_{k,diff}
\end{equation}

The key innovation of our approach lies in its selective slice matching strategy. Instead of a fixed window, we employ an adaptive search to identify the most plausible anatomical correspondence for each slice. First, we compute a similarity score for each potential slice pair:
\begin{equation}
S(k,j) = MMI(I_{LGE}^k, \phi_k(I_{T1m}^j)) + MMI(I_{LGE}^k, \phi_k(I_{T2m}^j))
\end{equation}

For the k-th LGE slice, we then identify the best-matching moving slice, $j_k^*$, by searching within a dynamically defined range. This range is constrained by the position of the previously matched slice ($j_{k-1}$) to maintain sequential order and ensure enough subsequent slices are available for future matches:
\begin{equation}
j_k^* = \argmin_{j \in [j_{k-1}, N-M+k]} S(k,j)
\end{equation}
\noindent where $j_{k-1}$ is the matching position of the previous slice, $N$ is the total number of moving image slices, and $M$ is the remaining number of slices from the current position k to the end of the LGE sequence. This ensures that:
\begin{equation}
\begin{cases}
j_k^* > j_{k-1} & \text{maintain sequence order} \\
j_k^* \leq N-M+k & \text{reserve sufficient slices}
\end{cases}
\end{equation}
This selective mechanism naturally handles through-plane motion while maintaining anatomical consistency within each 2D plane. The final registered volumes are reconstructed only from the reliably matched and transformed slice pairs:

\begin{equation}
V_{reg} = \{I_m^{j_k^*} \circ \phi_k^*\}_{k=1}^K
\end{equation}

This method enables our framework to focus on high-quality alignments while avoiding error propagation from forced matching of incompatible slices.

\begin{figure}[t!]
	\includegraphics[width=\textwidth]{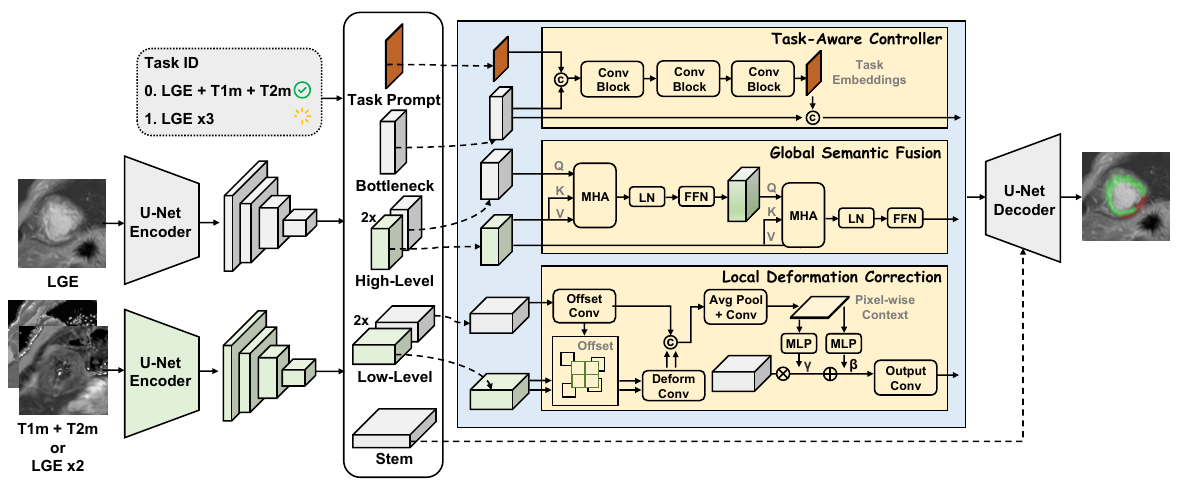}
	\centering
	\caption{Overview of the proposed HA-Net. The framework consists of dual encoders for multi-sequence CMR processing, with specialized alignment modules at different feature levels. Local deformation correction modules address spatial misalignment in low-level features, while cascaded cross-attention blocks enable semantic fusion at high-level features. A task-aware controller at the bottleneck adaptively modulates features based on different input images.} 
	\label{fig1}
\end{figure}  

\subsection{Hierarchical Alignment Network}
While SSA provides reliable inter-sequence alignment, minor local deformations and intensity variations may persist at the feature level. The low-level features primarily encode local tissue textures where geometric misalignments affect structural details, while high-level features capture semantic patterns where intensity variations become less significant. Therefore, we further propose a hierarchical alignment network (HA-Net) that progressively refines and fuses multi-sequence features at different semantic levels. Our network adopts a residual UNet backbone architecture with specialized fusion modules at different scales.

Let $F_{m}^l \in \mathbb{R}^{H_l \times W_l \times C_l}$ denote features output from encoder at level $l \in \{stem, low, high, bn\}$ for modality $m \in \{LGE, T1m, T2m\}$, where $H_l$, $W_l$, and $C_l$ represent spatial dimensions and channels. The network extracts initial features through the stem layer without cross-modal fusion, preserving modality-specific characteristics.

\noindent \textbf{Local Deformation Correction:} At low-level layers, we align features through deformable convolution and pixel-wise modulation. For each modality pair, the offset field $\Delta p$ and aligned features $F_{aligned}$ are computed as:

\begin{equation}
\begin{aligned}
\Delta p &= f_{offset}(F_{LGE}^{low}) \\
F_{aligned} &= \text{DeformConv}(F_{T1m/T2m}^{low}, \Delta p)
\end{aligned}
\end{equation}

The features are then modulated through learned parameters:

\begin{equation}
\begin{aligned}
[\gamma, \beta] &= f_{mod}(\text{GlobalPool}([F_{LGE}^{low}, F_{aligned}])) \\
F_{out}^{low} &= \gamma \cdot F_{LGE}^{low} + \beta
\end{aligned}
\end{equation}

\noindent where $f_{mod}$ maps the global context to modulation parameters $\gamma$ and $\beta$ for pixel-wise feature adjustment.

\noindent \textbf{Global Semantic Fusion:} High-level features undergo cascaded cross-attention for semantic alignment. The process consists of two sequential attention blocks:

\begin{equation}
\begin{aligned}
F_{1}^{high} &= \text{Attn}(Q=F_{LGE}^{high}, K,V=F_{T1m}^{high}) \\
F_{out}^{high} &= \text{Attn}(Q=F_{1}^{high}, K,V=F_{T2m}^{high})
\end{aligned}
\end{equation}

\noindent where the attention operation follows the standard scaled dot-product mechanism with multi-head projection.

\noindent \textbf{Task-Aware Controller:} Inspired by \cite{ye2023uniseg}, we incorporate a task-aware controller design that adaptively processes features based on different inputs. At the bottleneck, we integrate task information through a learned prompt $P_{task}$:

\begin{equation}
F_{out}^{bn} = \text{MLP}(\text{LayerNorm}([F_{LGE}^{bn}; P_{task}]))
\end{equation}

\noindent where $[\cdot;\cdot]$ represents channel-wise concatenation. This controller adapts the feature processing based on the input characteristics from different modalities.

\begin{table}[]
\centering
\caption{Quantitative evaluation results of different methods in myocardial lesion segmentation dataset. The best and second best results are highlighted in bold and underlined respectively. Myo: myocardium, ME: myocardial edema, MI: myocardial infarction.}
\label{tab1}
\resizebox{\columnwidth}{!}{%
\begin{tabular}{c|cc|cc|cc|cc}
\toprule
\multirow{2}{*}{Methods}     & \multicolumn{2}{c|}{Myo}                       & \multicolumn{2}{c|}{ME}                        & \multicolumn{2}{c|}{MI}                          & \multicolumn{2}{c}{Overall}                    \\
           & Dice (\%)             & HD95 (pixel)            & Dice (\%)              & HD95 (pixel)           & Dice (\%)              & HD95 (pixel)             & Dice (\%)              & HD95 (pixel)            \\ \midrule
nnU-Net \cite{isensee2021nnu}    & 76.46                 & 10.45                 & 62.68                  & 9.52                 & 45.30                  & {\ul 12.78}            & 61.48                  & {\ul 10.92}           \\
UNet++ \cite{zhou2018unet++}     & 77.51                 & {\ul 9.84}            & {\ul 65.03}            & 9.77                 & 44.26                  & 14.03                  & 62.27                  & 11.21                 \\
TransUNet \cite{chen2024transunet}  & 77.05                 & 9.87                  & 63.00                  & 11.08                & 43.25                  & 13.21                  & 61.10                  & 11.39                 \\
Swin UNETR \cite{hatamizadeh2021swin} & 75.23                 & 11.31                 & 61.39                  & 11.21                & 35.31                  & 14.94                  & 57.30                  & 12.49                 \\
UTNet \cite{gao2021utnet}      & 76.99                 & 10.25                 & 62.16                  & 10.15                & 42.06                  & 14.23                  & 60.40                  & 11.54                 \\
UMamba \cite{ma2024u}     & {\ul 77.58}           & 10.56                 & 64.93                  & \textbf{9.23$^\sim$} & {\ul 45.57}            & 13.11                  & {\ul 62.69}            & 10.97                 \\ \midrule
MFU-Net \cite{jiang2020max}    & 74.84                 & 10.89                 & 59.22                  & 12.53                & 36.68                  & 14.34                  & 56.91                  & 12.59                 \\
AWSNet \cite{wang2022awsnet}     & 76.50                 & 10.71                 & 63.06                  & 9.72                 & 40.00                  & 14.71                  & 59.85                  & 11.71                 \\
MyoPS-Net \cite{qiu2023myops}  & 77.20                 & 9.88                  & 63.90                  & 9.55                 & 44.15                  & 14.56                  & 61.75                  & 11.33                 \\
CAA-Seg (ours)    & \textbf{78.09$^\sim$} & \textbf{9.00$^\ddag$} & \textbf{65.49$^\ddag$} & {\ul 9.25}           & \textbf{51.11$^\ddag$} & \textbf{11.43$^\ddag$} & \textbf{64.89$^\ddag$} & \textbf{9.89$^\ddag$} \\ \midrule
\multicolumn{9}{l}{$\sim$:p-value>0.05, $\ddag$:p-value<0.05, in comparison between the best and second-best results.}                                                                                        \\ \bottomrule
\end{tabular}%
}
\end{table}

\section{Experiments and Results}

\textbf{Dataset and Experimental Setup:} We evaluate our method on an IRB-approved, large-scale in-house dataset comprising 397 patients with confirmed myocardial pathologies, randomly split into training (278 cases), validation (39 cases), and testing (80 cases) sets. Among these, all cases included annotations for myocardium (Myo) and myocardial edema (ME), while 208 cases presented with myocardial infarction (MI).

We use seven encoder stages with feature dimensions of \{32, 64, 128, 256, 512, 512, 512\} channels. The deformable convolution modules use 3×3 kernels with 18 sampling points. Input images are resampled and standardized to 384×384 pixels via crop or pad. The features naturally separate into low-level (64-256 channels) and high-level (512 channels) representations. The cross-attention blocks employ 8 attention heads with a hidden dimension of 256. The task controller generates a 128-dimensional task embedding that is projected to match feature dimensions.

Our comparisons include general medical segmentation frameworks (nnU-Net \cite{isensee2021nnu}, UNet++ \cite{zhou2018unet++}, TransUNet \cite{chen2024transunet}, Swin UNETR \cite{hatamizadeh2021swin}, UTNet \cite{gao2021utnet}, UMamba \cite{ma2024u}) and cardiac-specific approaches (MFU-Net \cite{jiang2020max}, AWSNet \cite{wang2022awsnet}, MyoPS-Net \cite{qiu2023myops}). General frameworks were evaluated using only LGE images due to performance degradation with multi-sequence input, while cardiac-specific approaches utilized multi-sequence data with their dedicated fusion mechanisms. For fair comparison, all methods follow identical preprocessing and postprocessing protocols as nnU-Net, and are trained for 200 epochs with an initial learning rate of 0.01 using cosine decay scheduler and batch size of 16 on an NVIDIA RTX 4090 GPU. The network was trained using a combination of Dice loss and Cross-Entropy loss, consistent with the standard nnU-Net framework. The results are assessed using Dice coefficient (\%) and 95\% Hausdorff Distance (HD95, in pixels) across three targets: Myo, ME, and MI. Statistical significance is assessed using the Wilcoxon signed rank test between our method and other approaches.

\begin{figure}[t!]
	\includegraphics[width=0.9\textwidth]{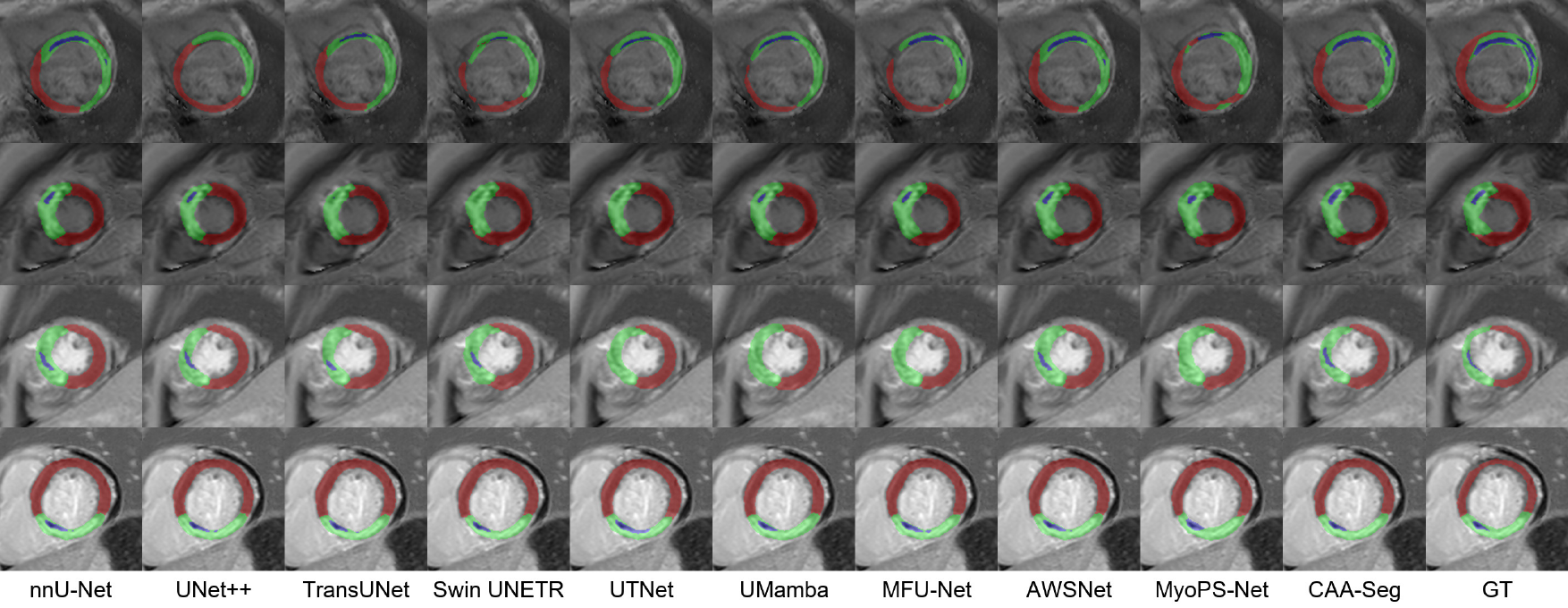}
	\centering
	\caption{Visual comparison of myocardial lesion segmentation results across different methods with myocardium shown in red, myocardial edema in green, and myocardial infarction in blue.} 
	\label{fig3}
\end{figure}

\begin{table}[]
\centering
\caption{Ablation study of CAA-Seg, analyzing the impact of different registration methods, network architectures, and input sequences on segmentation performance.}
\label{tab2}
\resizebox{0.7\columnwidth}{!}{%
\begin{tabular}{c|ccc|cc|cc|cc}
\toprule
\multirow{2}{*}{Settings} & \multicolumn{3}{c|}{Registration}    & \multicolumn{2}{c|}{Network} & \multicolumn{2}{c|}{Input} & \multicolumn{2}{c}{Overall}    \\
                          & Baseline   & MvMM \cite{zhuang2018multivariate}       & SSA        & nnU-Net       & HA-Net       & LGE         & Multi            & Dice (\%)       & HD95 (pixel)  \\ \midrule
1                         & \checkmark &            &            & \checkmark    &              &                  & \checkmark       & 59.63          & 12.17         \\
2                         & \checkmark &            &            &               & \checkmark   &                  & \checkmark       & 60.12          & 11.92         \\
3                         &            & \checkmark &            &               & \checkmark   &                  & \checkmark       & {\ul 62.98}    & {\ul 10.83}   \\
4                         &            &            & \checkmark &               & \checkmark   & \checkmark       &                  & 61.51          & 11.85         \\
5 (proposed)                         &            &            & \checkmark &               & \checkmark   &                  & \checkmark       & \textbf{64.89} & \textbf{9.89} \\ \hline
\end{tabular}%
}
\end{table}

\textbf{Results:} The experimental results comparing different methods for myocardial pathology segmentation are presented in Table \ref{tab1}. Our CAA-Seg achieves consistent improvements across all metrics, with notable gains in the challenging MI segmentation task where it reaches a Dice score of 51.11\% (p\textless 0.05 improvement over the second-best result). The performance advantage is particularly evident in overall metrics, where CAA-Seg obtains a mean Dice of 64.89\% and HD of 9.89 pixels, representing statistically significant improvements over previous methods. Interestingly, while UMamba shows competitive performance in Myo and ME segmentation, its effectiveness diminishes in MI detection, highlighting the advantages of our alignment-aware design. In terms of computational efficiency, CAA-Seg remains competitive. The average inference time per case was 0.71s with 1.8GB of GPU memory, compared to nnU-Net (0.52s, 1.1GB) and UMamba (0.62s, 1.3GB). While slightly more intensive due to the explicit alignment, our framework offers substantial performance gains for a modest increase in computational cost.

The qualitative results visualized in Fig. \ref{fig3} further demonstrate how our method effectively handles complex cases where traditional approaches struggle with multi-sequence alignment and feature fusion. CAA-Seg achieves more accurate target localization, especially in regions with minor pathological changes that are typically challenging to identify. 

\textbf{Ablation Study:} We further conduct ablation experiments to validate the effectiveness of each component in our framework. The results in Table \ref{tab2} demonstrate that our SSA strategy combined with the HA-Net achieves the best performance, significantly outperforming both traditional registration methods and single-modality approaches. Notably, while the multivariate mixture model (MvMM) \cite{zhuang2018multivariate} registration shows competitive results, the integration of SSA with multi-sequence input further improves performance by handling spatial misalignment more effectively.

\section{Discussion and Conclusion}
In this work, we presented CAA-Seg, a composite alignment-aware framework for myocardial lesion segmentation in multi-sequence CMR images. Our method effectively handles the heterogeneous nature of multi-sequence cardiac imaging data, where misalignments stem from both acquisition protocols and tissue characteristics. The selective slice alignment strategy successfully establishes reliable correspondence between LGE and mapping sequences, while the hierarchical alignment network enables effective feature fusion across modalities with varying intensity distributions and spatial resolutions. Experimental validation on a large-scale dataset of 397 cases demonstrates significant performance improvements across all lesion regions. 

We acknowledge that the performance of CAA-Seg may degrade in cases with extreme inter-sequence misalignment beyond the capture range of our alignment module, or for very small and indistinct lesions that are challenging even for expert analysis. Future work could focus on extending the framework to handle more complex clinical scenarios and incorporating temporal information from cine sequences. The composite alignment strategy demonstrated in this work shows promise for application in other multi-sequence medical imaging tasks that require cross-modality feature fusion.

\begin{credits}
\subsubsection{\discintname}
The authors have no competing interests.
\end{credits}

\bibliography{mybibliography}

\end{document}